\documentclass[hyper]{JHEP} 

\usepackage{epsfig}

\newcommand\fverb{\setbox\pippobox=\hbox\bgroup\verb}
\newcommand\fverbdo{\egroup\medskip\noindent%
			\fbox{\unhbox\pippobox}\ }
\newcommand\fverbit{\egroup\item[\fbox{\unhbox\pippobox}]}
\newbox\pippobox
\title{Exact Solutions of Open Bosonic
 String Field Theory}
\author{ J. Kluso\v{n}
\footnote{On leave from Masaryk University, Brno}\\
Institutionen f\"or teoretisk fysik\\
BOX 803, SE- 751 08 
Uppsala, Sweden \\
E-mail: \email{josef.kluson@teorfys.uu.se}}

\preprint{\hepth{0202045}}

\abstract{In this paper we present two solutions
of open bosonic string field theory defined on 
D0-brane background  that correspond to 
the change of the  D0-brane position
 in the transverse space.}

\begin{document}
%%%%%%%%%%%%%%%%%%%%%
%%%%Introduction %%%%%%%%%
%%%%%%%%%%%%%%%%%%%%
\section{Introduction}\label{first}
During last few years many different solutions have
been studied in string field theory
\cite{WittenSFT}. These solutions
include tachyon vacuum and lump solutions 
(For review and extensive list of references,
see \cite{Ohmori,Desmet,BelovR}.) and solutions
representing marginal deformations
\cite{SenM1,SenM2}. Other remarkable solutions
have been studied recently in papers
\cite{Takahashi1,Takahashi2}\footnote{As we will see the
solution given in \cite{Takahashi1} has
a lot of common with the solution given in
this paper.}. And recently there
were also many papers published regarding to the
string field theory formulated around closed
string vacuum (See, for example
\cite{RSZF,rsz-3,GT01,GT02,rsz-4,rsz-5,0108150,
MooreTaylor,0111092,rsz-6,0201095,AGM,Marino,AGM2}.)

In our recent paper \cite{KlusonGS} we have suggested
possible exact solutions of the Berkovits
superstring field theory
\cite{Berkovits1,Berkovits2,BerkovitsR}. Then
we have extended our approach to the open bosonic
string field theory \cite{WittenSFT} as 
well.
In this paper we will continue our 
study. We present two solutions that correspond
to the marginal deformation of the theory
describing single D0-brane. The first solution
corresponds to the translation in the target
space orthogonal to D0-brane. Since we
presume that our D0-brane lives in flat space-time
its position should be irrelevant for its
description. On the other hand, since D0-brane
can be regarded as a defect that breaks 
Poincare invariance, we will see that there 
exist nontrivial solution. Our strategy is as
follows. We use the fact that closed string theory
is invariant under Poincare transformations
which are broken by boundary conditions
in the open string case describing D0-brane).
 Then we can construct generator of the translation
that does not commute with the open string
Hamiltonian thanks to Dirichlet boundary
conditions, but can serve as a generator of
translation that shifts position of
D0-brane. We will see that this is really this
case. Following our general discussion given
in \cite{KlusonGS} and reviewed in section
(\ref{second}) we will construct solution of the
open string field theory corresponding to
this marginal deformation. We will see
that in case of small marginal parameter
 the  BRST operator changes exactly in
the same way as we can expect from CFT analysis
with accord with the very nice analysis performed
in \cite{SenSFT}. 
Then we will discuss possible form of the 
string  field redefinition
and we will show that after this redefinition
 fluctuation fields are 
the same as the original one as a 
 consequence of the invariance
of our configuration under translation. 

Then we will extend our solution to the
case of finite marginal deformation. We will
see that in this case the BRST operator obtains
additional term which is not expected from
the CFT analysis \cite{Schomerus}. At present
we do not understand why such a term emerges.
However we can still argue that thanks to the
form of the  new BRST operator we 
can perform such a field
redefinition that maps the new  BRST operator 
to the original one and string field describing
the fluctuation around new solution has the same
form as the original one thanks to the translation
invariance of this configuration.

As the second example we present rotation of the
D0-brane in the  target space. As in previous
case, Dirichlet boundary conditions break this
rotation invariance hence we can expect that
the generator of rotation does not commute with 
the BRST operator and hence we can get
nontrivial solution.   We will also discuss the problem
of the field redefinition and we will argue that
in this case the field changes since generator
of rotation acts non trivially 
 on  tensor indexes of
various string states, however thanks to the invariance of
the bulk part of the BRST operator under 
rotation we will argue that the physical spectrum
is the same. Again this can be interpreted
 as a trivial
consequence  of the isotropy of the target space.

In conclusion we outline our results and 
suggest possible extension of our research. 
%%%%%%%%%%%%%%%%%%%%
%%%Marginal %%%%
%%%%%%%%%%%%%%%%%%%%
\section{Solutions of open bosonic
string field theory}\label{second}
As we claimed in the introduction we restrict in
this paper to the case of the bosonic open string field theory
\cite{WittenSFT} since generalization to the
Berkovits' superstring field theory
\cite{Berkovits1,Berkovits2,BerkovitsR} is straightforward.
This section mainly reviews results given in our
previous paper \cite{KlusonGS}.

The string field theory (SFT) action has a form
\footnote{For very nice recent reviews of
string field theory, see \cite{Ohmori,Desmet,BelovR}.}
\begin{equation}\label{SFTaction}
S=\frac{1}{2}\int \Phi\star Q (\Phi)+
\frac{1}{3}\int \Phi\star \Phi\star
\Phi \ .
\end{equation}
Variation of  the action leads to the equation of motion
\begin{equation}\label{eqm}
Q(\Phi)+\Phi\star \Phi=0 \ .
\end{equation}
According to the standard procedure let us consider the
expansion of the string field around any solution
of the equation of motion (\ref{eqm}) $\Phi_0$
\begin{equation}\label{gfield}
\Phi=\Phi_0+\Psi \ .
\end{equation}
After inserting (\ref{gfield}) into (\ref{SFTaction}) 
and using (\ref{eqm}) we obtain SFT action
for $\Psi$ that has the same form as the original
one (\ref{SFTaction})
\begin{equation}
S=\frac{1}{2}\int \Psi\star \tilde{Q} (\Psi)+
\frac{1}{3}\int \Psi\star \Psi\star
\Psi \ 
\end{equation}
however with the modified BRST operator
\begin{equation}
\tilde{Q}(X)=
Q(X)+\Phi_0\star X-(-1)^{|X|} X\star \Phi_0 \ ,
\end{equation}
where $X$ is any string field of ghost number $|X|$. 
Let us now presume that the  new BRST operator
$\tilde{Q}$ can be written as 
\cite{KlusonGS}
\begin{equation}\label{newBRSTK}
\tilde{Q}(X)=e^{-K}\left( Q(e^{K}(X))\right) \ ,
\end{equation}
where $K$ is an  operator of ghost charge  equal to zero
that obeys following rules  \cite{F2,rsz-5}
\begin{eqnarray}\label{Krule}
K(X\star  Y)=K(X)\star Y+
X\star K(Y) , \ \nonumber \\
\int K(X)\star Y=-\int X  
\star K(Y) \ .
\nonumber \\
\end{eqnarray}
In order to have a well defined string field theory
for fluctuation field $\Psi$, 
$\tilde{Q}$ must obey following axioms
\cite{WittenSFT}
\begin{eqnarray}\label{ax}
\tilde{Q}^2=0 \  , \nonumber \\
\tilde{Q}(X\star Y)=\tilde{Q}(X)
\star Y+(-1)^{|X|}X\star \tilde{Q}(Y) \ ,
\forall X, Y \ , 
\nonumber \\
\int \tilde{Q}(X)=0 \ , \forall X \ , 
\nonumber \\
\end{eqnarray}
where $X, Y$ are arbitrary string fields
of ghost numbers $|X|$ and $|Y|$ respectively.  
 Firstly, we can
easily show that
\begin{equation}
\tilde{Q}^2=e^{-K}Qe^{K}
e^{-K}Qe^{K}=e^{-K}Q^2
e^{K}=0 \ 
\end{equation}
since the original BRST operator $Q$ obeys 
(\ref{ax}) by definition. 
 We also have
\begin{eqnarray}
\tilde{Q}(X \star Y)=
e^{-K}\left[Q\left(e^{K}(X\star Y)\right)\right]=
e^{-K}\left[Q\left(e^{K}(X)\star e^{K}(Y)\right)
\right]=
\nonumber \\
=e^{-K}\left[Q\left(e^{K}(X)\right)\star e^{K}(Y)+
(-1)^{|X|}
e^{K}(X)\star Q\left(e^{K}(Y)\right)\right]=\nonumber \\
=e^{-K}\left[Q\left(e^{K}(X)\right)\right]\star
e^{-K}e^{K}(Y)+(-1)^{|X|}e^{-K}
e^{K}(X)e^{-K}\left[Q\left(e^{K}(Y)\right)\right]=
\nonumber \\
=\tilde{Q}(X)\star Y
+(-1)^{|X|}X\star \tilde{Q}(Y) \ , \nonumber \\
\end{eqnarray}
where we have used
\begin{equation}
e^{K}(X \star Y)=e^{K}(X)\star e^{K}(Y)
 , \ 
e^{-K}(X\star Y)=e^{-K}(X)\star e^{-K}(Y) \ ,
\end{equation}
which follows from (\ref{Krule}). 
 We can also easily see
that (\ref{newBRSTK}) obeys the third axiom
in (\ref{ax}) 
\begin{equation}\label{Q0}
\int \tilde{Q}(X)=
\int  e^{-K}\left[Q\left(e^{K}(X)\right)\right]\star\mathcal{I}=
\int Q\left(e^{K}(X)\right)\star e^K(\mathcal{I})=
\int Q\left(e^{K}(X)\right)=0 
 \ ,
\end{equation}
where $\mathcal{I}$ is "identity", the ghost number 
 zero field (For recent discussion of
some properties of this field, see \cite{Feng1,Ohmori1}.)
which is defined as
\begin{equation}
\mathcal{I}\star X=X\star \mathcal{I}=X \ , \forall X \ .
\end{equation}
Then we  have 
\begin{equation}
K(X)=K(\mathcal{I}\star X)=
K(\mathcal{I})\star X+\mathcal{I}\star K(X)
=K(X) \Rightarrow K(\mathcal{I})=0
\end{equation}
and consequently 
\begin{equation}
e^{-K}(\mathcal{I})=
e^K(\mathcal{I})=\mathcal{I} \ .
\end{equation} 

We have  shown  that the new BRST operator 
$\tilde{Q}$ given in (\ref{newBRSTK}) obeys all
axioms (\ref{ax}). Now we must find such a solution
of the equation of motion (\ref{eqm}) that leads to
this proposed form of the BRST operator (\ref{newBRSTK}). 
It  is easy  to see that the field
\cite{KlusonGS}
\begin{equation}\label{sol}
\Phi_0=e^{-K_L(\mathcal{I})}\star Q
(e^{K_L(\mathcal{I})})
\end{equation}
 is desired  solution of the equation of motion
(\ref{eqm})
since we have
\begin{equation}
Q  (\Phi_0)=Q(e^{-K_L(\mathcal{I})}
\star Q(e^{K_L(\mathcal{I})}))=
-e^{-K_L(\mathcal{I})}
\star Q(e^{K_L(\mathcal{I})})\star 
e^{-K_L(\mathcal{I})}
\star   Q(e^{K_L(\mathcal{I})})
\end{equation}
and
\begin{equation}
\Phi_0\star \Phi_0=e^{-K_L(\mathcal{I})}
\star Q(e^{K_L(\mathcal{I})})
\star e^{-K_L(\mathcal{I})}
\star Q(e^{K_L(\mathcal{I})}) \ .
\end{equation}
with $K$  obeying (\ref{Krule})
and where  $K$ is split in two parts
according to  \cite{Horowitz}
$K=K_L+K_R$.
It can be shown \cite{KlusonGS} that the
solution (\ref{sol}) leads to the BRST operator
(\ref{newBRSTK}). As was argued in
\cite{KlusonGS} we can perform
field redefinition 
\begin{equation}
\Psi=e^{-K}(\tilde{\Psi}) \ 
\end{equation}
that leads to the string field theory action
for fluctuation field $\tilde{\Psi}$ with
the BRST operator of the same form as the original
one $Q$
and the string  field $\tilde{\Psi}$
 now describes open string theory defined
around  the new background configuration (For more
detailed and extensive discussion of the 
string redefinition in  the string
field theory we again recommend very nice
and  seminal paper
\cite{SenSFT}.).

In  the next two sections we will continue our
study and give two examples of the solutions
reviewed above.
%%%%%%%%%%%%%%%%%%%%%%%%%%%%%%%%%%
%%%%%%%Examples of K %%%%%%%%%%%%%%%%%%
%%%%%%%%%%%%%%%%%%%%%%%%%%%%%%%%%
\section{Translation of D0-brane}\label{third}
In this section we will describe solution
of the string field theory equation of motion 
corresponding to the small translation in the
target space. We will see that for small value of the
marginal parameter $a^I, I=1,\dots,25
$ we obtain such a form
of the deformed BRST operator  as we will get
from the perturbation of the 
open string Lagrangian   with the
marginal operator. 
According to the general method
outlined in the previous section
we will  perform string field theory redefinition
that maps the new BRST operator to the original
one and the old string field to the new string
field describing fluctuation around  new
configuration. However thanks to the fact that our solution
corresponds to the translation of a single D0-brane
in the target space, it is clear that the Hilbert
spaces of the original and final configuration
must be isomorphic. We will confirm explicitly
this conclusion very easily using the
fact that the generator of the translation
in the target space acts trivially on the string
state describing fluctuation of D0-brane since
these states carry zero momentum in the transverse
direction. Then in the next section we will construct
solution corresponding to the  finite translation where
we will see some subtle facts that are not completely
clear to us at present. 

For the beginning we will  review well
known facts about conserved charges in
the bosonic  string theory, following
\cite{WittenSB}. 
Let us consider an  action for the free string
 \cite{WittenSB}
\begin{equation}\label{stringaction}
S=-\frac{1}{4\pi\alpha'}\int 
d^2\sigma \eta^{\alpha\beta}\partial_{\alpha}
X^{\mu}\partial_{\beta}X^{\nu}\eta_{\mu\nu } \ ,
\end{equation}
where $\eta^{\alpha\beta}$
is two-dimensional Minkowski metric with
signature $\eta_{\alpha\beta}=diag(-1,1)$  and $\eta_{\mu\nu}$
is 26-dimensional Minkowski metric in the target space-time
with signature $(-,+,\dots,+)$. 
Let us suppose that this action describes string ending
on D0-brane. In other words, on the boundary of the world-sheet
we have Neumann boundary condition for $X^0$ and
Dirichlet boundary condition for $X^I, I=1.\dots, 25$, where the
 boundary of the world-sheet is at the points
 $\sigma=0,\pi$. Now it is easy to
see that this boundary conditions break Poincare symmetry under
translation, which from the point of view of two dimensional 
world-sheet theory is  global symmetry that acts on the string
coordinate $X^{\mu}(\sigma)$ as
\begin{equation}
\delta X^{\mu}=\epsilon^{\mu} \ .
\end{equation}
Since D0-brane is placed 
 in the point $X^I=0$ it is clear that the variation
of $X^I$  is equal to zero $\delta X^I=0$ on the boundary
of the world-sheet which 
 leads to the breaking of the Poincare invariance.
This seems to be
 natural fact  since D0-brane can be regarded
as a defect in the target space. 
 As a result, we cannot define conserved
generator of translation  $P_I$ that
commutes with the  world-sheet Hamiltonian. 
On the other hand, $P_I$ is the symmetry generator
of closed string theory and can be determined using Noether's
method \cite{WittenSB} as follows. 
Since the string action (\ref{stringaction})
is invariant under variation $\delta X^{\mu}=\epsilon^{\mu}$ we
now presume that $\epsilon$ is a function of $\sigma, \tau$. Then we
have
\begin{eqnarray}
\delta S=-\frac{1}{2\pi\alpha'}\int d^2\sigma \eta^{\alpha\beta}
\partial_{\alpha}X^{\mu}\partial_{\beta}\delta X^{\nu}\eta_{\mu\nu}=
-\frac{1}{2\pi\alpha'}\int d^2\sigma \eta^{\alpha\beta}
\partial_{\alpha}X^{\mu}\partial_{\beta}\epsilon^{\nu}\eta_{\mu\nu}
\Rightarrow\nonumber \\
\Rightarrow \partial_{\alpha}\left(\frac{1}{
2\pi\alpha'}\eta^{\alpha\beta}\partial_{\beta}X^{\mu}
\eta_{\mu\nu}\right)=0=
\partial_{\alpha}(\eta^{\alpha\beta}P_{\beta\nu})=0 \ ,
P_{\beta\mu}=\frac{1}{2\pi\alpha'}
\partial_{\beta}X^{\nu}\eta_{\nu\mu} \ . \nonumber \\
\end{eqnarray}
Then we have  conserved quantity (In case of closed
string or an  open string obeying
Neumann boundary conditions.)
\begin{equation}
P_{\mu}=i\int_0^{\pi}d\sigma P_{0\mu}=
\frac{i}{2\pi\alpha'}\int_0^{\pi} d\sigma 
\dot{X}^{\nu}\eta_{\nu\mu} \ , \partial_{\tau}X=\dot{X} , 
\partial_{\sigma}X=X' \ .
\end{equation}
Since $X^{\mu}(\sigma)$ is canonical coordinate we obtain
 from the action (\ref{stringaction}) its canonical
 conjugate momentum
\begin{equation}
P_{0\mu}(\sigma)=\frac{\delta L}
{\delta \dot X^{\mu}(\sigma)}=
\frac{1}{2\pi\alpha'}
\dot X^{\nu}(\sigma)\eta_{\nu\mu} 
\end{equation}
with the following commutation relation
\begin{equation}\label{com}
[P_{0I}(\sigma),X^{J}(\sigma')]=
-i\delta_D (\sigma,\sigma') \delta_{I}^{J} \ ,
\end{equation}
where $\delta_D (\sigma,\sigma')$ is a delta function
that  obeys Dirichlet
boundary conditions so that can be written as
\footnote{More information about delta function
can be found in Appendix.}
\begin{equation}
\delta (\sigma,\sigma')=\frac{1}{\pi}
\sum_{n=-\infty}^{\infty}\sin n\sigma
\sin n \sigma'\ .
\end{equation}
Using (\ref{com}) we can calculate commutator
\begin{eqnarray}
[H,P_I]=\left[\frac{1}{4\pi\alpha'}\int_0^{\pi}
d\sigma ( (2\pi\alpha')^2 P_{0}^2(\sigma)+
X'(\sigma)^2), i\int_0^{\pi}d\sigma'
P_{0I}(\sigma')\right]=
\nonumber \\
=-\frac{1}{2\pi\alpha'}\int_0^{\pi}
d\sigma  \int_0^{\pi}d\sigma'
\partial_{\sigma}\delta_D(\sigma,\sigma')
\delta_I^K X'_K(\sigma)=
\frac{1}{4\pi\alpha'}\int_0^{\pi}
d\sigma (\delta_N(\sigma,\pi)-
\delta_N(\sigma,0))X'_I(\sigma) \ , \nonumber \\
\end{eqnarray}
 where we have used (\ref{dDN}). We see that
in the case of the string obeying Dirichlet boundary
conditions the upper commutator is nonzero
which is a consequence of the breaking of Poincare invariance
with D0-brane background. On the other 
hand, 
open string with Neumann boundary conditions obeys
$\left.X'(\sigma)\right|_0^{\pi}=0$ so that
the upper commutator is equal to zero 
and hence there is conserved charge $P_{\mu}$.

Let us consider   $K$ in the form
\begin{equation}\label{Kdef}
K=a^{\mu}P_{\mu}=
\frac{ai}{2\pi\alpha'}\int_0^{\pi}
d\sigma \dot{X}^1 \ , a^{\mu}=(0,a,0,\dots,0) \ . 
\end{equation}
It is easy to see, using contour
arguments as in \cite{Ohmori,Horowitz}
 that $K$ given in
(\ref{Kdef}) obeys all conditions (\ref{Krule})
We can also prove that $K$ obeys (\ref{Krule}) using
the fact that its action on any string
field $X$ can be defined as
follows
\begin{equation}
K(X)=K_L(\mathcal{I})\star X-X\star K_R(
\mathcal{I})=\mathcal{I}\star K_R(X)+
K_L(X)\star \mathcal{I}=K(X)
\end{equation}
and   operators of this form, as was
reviewed recently in \cite{rsz-5} 
and can be proved  very easily,
obey  (\ref{Krule}).

Let  us start with  situation when $a<<1$ so that we have
\begin{equation}\label{phi0}
\Phi_0=e^{-K_L(\mathcal{I})}\star Q
(e^{K_L(\mathcal{I})})\sim
(\mathcal{I}-K_L(\mathcal{I}))\star Q
(\mathcal{I}+K_L(\mathcal{I}))=
Q(K_L(\mathcal{I})) \ .
\end{equation}
We can write
(\ref{phi0})  as
\begin{eqnarray}
\Phi_0=Q(K_L(\mathcal{I}))=Q_L(K_L(\mathcal{I}))+
Q_R(K_L(\mathcal{I}))=\nonumber \\
=Q_L(K_L(\mathcal{I}))+K_L(Q_R(\mathcal{I}))=
Q_L(K_L(\mathcal{I}))-K_L(Q_L(\mathcal{I}))=
[Q_L,K_L](\mathcal{I}) \  \nonumber \\
\end{eqnarray}
using \cite{Horowitz}
\begin{equation}
Q_L(\mathcal{I})=-Q_R(\mathcal{I}) \ 
\end{equation}
and 
\begin{eqnarray}
Q_R(K_L(\mathcal{I}))=\mathcal{I}\star
Q_R(K_L(\mathcal{I}))=-Q_L(\mathcal{I})\star
K_L(\mathcal{I})=\nonumber \\
=Q_L(\mathcal{I})\star K_R(\mathcal{I})=
-K_L(Q_L(\mathcal{I}))\star \mathcal{I}=
-K_L(Q_L(\mathcal{I})) \ .
\nonumber \\
\end{eqnarray}
Let us define
\begin{equation}
\Phi_0=D_L(\mathcal{I})=[Q_L,K_L](\mathcal{I}) \ .
\end{equation}
Then the new BRST operator $Q'$ 
has the form
\begin{equation}\label{Q'}
 Q'(X)=
 Q(X)+ D_L(\mathcal{I})\star
X -(-1)^{|X|}
X \star D_L(\mathcal{I})=[Q-D](X)
\end{equation}
using 
\begin{eqnarray}
               D_L(\mathcal{I})\star
X=-\mathcal{I}
\star D_R(X)= -D_R(X) \ , \nonumber \\
-(-1)^{|X|}X \star D_L(\mathcal{I})=
(-1)^{|X|}X\star D_R(\mathcal{I})=-
 D_L(X) , \nonumber \\
D_L(X)\star Y=-(-1)^XX\star D_R(Y) \ .\nonumber \\
\end{eqnarray}
The last formula can be proven in the same
way  as in
\cite{Horowitz}.
Let us calculate the commutator $[Q,K]$.
 Since we have
\begin{eqnarray}
Q=\frac{1}{\pi}\int_0^{\pi}d\sigma J_{0}(\sigma)=
\frac{1}{\pi}\int_0^{\pi}
d\sigma c^{\alpha}(\sigma)T_{\alpha 0}(\sigma)+Q_{ghost}
=\nonumber \\
=\frac{1}{\pi}\int_0^{\pi}d\sigma 
\left[c^0\frac{1}{2}\left(\dot{X}^2+X'^2\right)+
c^1\dot{X}X'\right]+Q_{qhost} \ ,\nonumber \\
\end{eqnarray}
where $Q_{qhost}$ denotes the pure ghost part of 
the BRST operator which is not important
for our calculation 
and $T_{0\alpha}$ is the matter
stress energy tensor. We see that
we  must calculate following commutators
\begin{equation}
[T_{00}(\sigma),P(\sigma')]=
\frac{1}{2}[X'^2(\sigma),P(\sigma')]=i\partial_{\sigma}
\delta_D (\sigma,\sigma')X'(\sigma) \ ,
\end{equation}
\begin{equation}
[T_{10}(\sigma),P(\sigma')]=
[\dot{X}(\sigma)X'(\sigma),P(\sigma')]=
i\dot{X}(\sigma)\partial_{\sigma}\delta_D
(\sigma,\sigma')\ ,
\end{equation}
where we have used the notation
\begin{equation}
P_{01}(\sigma)=P(\sigma), \
X^1(\sigma)=X(\sigma) \ .
\end{equation}
Then we can write
\begin{eqnarray}\label{Dc}
D=[Q,K]=\frac{ai}{\pi}\int_0^{\pi}d\sigma
\int_0^{\pi}d\sigma'[J_{0}(\sigma),P(\sigma')]=
\nonumber \\
=\frac{ai}{\pi}\int_0^{\pi}d\sigma
\int_0^{\pi}d\sigma'\left[
c^0(\sigma)[T_{00}(\sigma),P(\sigma')]+
c^1(\sigma)[T_{10}(\sigma),P(\sigma')]\right]=
\nonumber \\
=\frac{ai}{\pi}\int_0^{\pi}d\sigma
\int_0^{\pi}d\sigma'\left[
c^0(\sigma)i\partial_{\sigma}\delta_D(\sigma,\sigma')
X'(\sigma)+ic^1(\sigma)\partial_{\sigma}
\delta_D(\sigma,\sigma')
\dot{X}(\sigma)\right]=\nonumber \\
=\frac{a}{\pi}\int_0^{\pi}d\sigma
\left[c^0(\sigma)X'(\sigma)(\delta_N (\sigma,\pi)
-\delta(\sigma,0))
+c^1(\sigma)\dot{X}(\sigma)(
\delta_N(\sigma,\pi)-\delta_N(\sigma,0))
\right]=\nonumber \\
=\frac{a}{\pi}\int_0^{\pi}d\sigma
 \left[ \left(c^+\partial_+X-
c^-\partial_-X\right)\left(
\delta_N(\sigma,\pi)-
\delta_N (\sigma,0)\right)\right] \ ,
\nonumber \\
\end{eqnarray}
using (\ref{dDN}) and also 
where we have used the notation
\begin{eqnarray}
\sigma^+=\tau+\sigma, \ \sigma^-=
\tau-\sigma \ , \nonumber \\
 \partial_+=
\frac{\partial}{\partial\sigma^+}=\frac{1}{2}
(\frac{\partial}{\partial \tau}+
\frac{\partial}{\partial \sigma}), \
\partial_-=\frac{\partial}
{\partial \sigma^-}=\frac{1}{2}(
\frac{\partial}{\partial \tau}-
\frac{\partial}{\partial \sigma}) \ , 
\nonumber \\
c^+=c^0+c^1 \ , c^-=c^0-c^1 \ ,
\nonumber \\
c^+\partial_+X-c^-\partial_-X=
\frac{1}{2}\left[(c^0+c^1)(\dot{X}+X')-
(c^0-c^1)(\dot{X}-X')\right]=\nonumber \\
=\frac{1}{2}[c^0\dot{X}+c^0X'+
c^1\dot{X}+c^1 X'-c^0\dot{X}+
c^0X'+c^1\dot{X}-c^1X']
=c^0X'+c^1 \dot{X} \ . \nonumber \\
\end{eqnarray}
From (\ref{Q'}) and
 (\ref{Dc})  we can conclude that our solution leads
to the deformation of the stress energy tensor
\footnote{Similar construction appeared in case of relevant
perturbation in \cite{Zabzine}.}
\begin{equation}
T'_{++}=T_{++}-\partial_+X(\delta_N(\sigma,\pi)-
\delta_N(\sigma,0)) \ ,
T'_{--}=T_{--}+\partial_-X(\delta_N(\sigma,\pi)
-\delta_N(\sigma,0)) \ ,
\end{equation}
using
\begin{eqnarray}
T_{++}=\frac{1}{2}(T_{00}+T_{10}) \ ,
T_{--}=\frac{1}{2}(T_{00}-T_{10}) \ ,
\nonumber \\
c^+T_{++}+c^-T_{--}=
c^0T_{00}+c^1T_{10} \ . \nonumber \\
\end{eqnarray}
In other words, our solution corresponds to the marginal deformation
of the world-sheet theory in  precise the
same way as in \cite{SenSFT}.

Now we can ask the question of the field redefinition. Recall, that
according to our general discussion in
section (\ref{second}) we know that after suitable field
redefinition the new  BRST operator $Q'$ is mapped to
the original BRST operator  $Q$
and  the new string field is  given
as
\begin{equation}
\tilde{\Psi}=e^K(\Psi) \ .
\end{equation} 
The operator $K$ acts trivially 
on all original fields $\Psi$
since  all string states corresponding
to the string living on D0-brane  do not carry
momentum in the
 transverse directions and hence are
invariant under the translation. This is the result that
we can expect since our solution should not
depend on one particular position of D0-brane
in the transverse space. 
However as we will see on the solution
corresponding to the  rotation that 
open  string states  transform non trivially
as a consequence of their nontrivial  tensor structure.

Now let us turn to the problem of the finite translation
using our general prescription given in 
the (\ref{second}) section. 
%%%%%%%%%%%%%%%%%%%%%%%%%%%%%
%%Finite translation %%%%%%%
%%%%%%%%%%%%%%%%%%%%%%%%%%%%%
\section{Finite translation}\label{fourth}
Let us consider the case of finite deformation where
 the solution is given as
\begin{equation}
\Phi_0=e^{-K_L(\mathcal{I})}\star Q(e^{K_L(
\mathcal{I})}) \ .
\end{equation}
In order 
to calculate $\Phi_0$ we define following function 
\cite{KlusonNSFT2}
\begin{equation}
F(t)=e^{-K_L(\mathcal{I})t}\star Q(e^{K_L(
\mathcal{I})t}) \ , \Phi_0=F(1) \ 
\end{equation}
so that
\begin{equation}
F(1)=F(0)+\sum_{n=1}^{\infty}
\frac{1}{n!}\frac{d^nF(0)}{d^nt} \ ,
\end{equation} 
where
 \cite{KlusonNSFT2}
\begin{eqnarray}\label{F}
\frac{d^nF(t)}{dt^n}=
e^{-tK_L(\mathcal{I})}\star[[Q(K_L(\mathcal{I})),
K_L(\mathcal{I})],\dots,K_L(\mathcal{I})]\star
e^{K_L(\mathcal{I})t} , n>1 ,
\nonumber \\
F(0)=Q(\mathcal{I})=0 \ ,
\frac{dF(0)}{dt}=Q(K_L(\mathcal{I}))=
[Q,K]_L(\mathcal{I}) \ . \nonumber \\
\end{eqnarray}
 We see that 
we must calculate commutators given in
(\ref{F}). The first commutator was calculated 
in the (\ref{third}) section
 with the result (\ref{Dc})
\begin{equation}
[Q,K]=\frac{a}{\pi}\int_0^{\pi}
d\sigma \left[
 \left(c^+\partial_+X-
c^-\partial_-X\right)\left(
\delta_N(\sigma,\pi)-
\delta_N (\sigma,0)\right)
\right] \ .
\end{equation}
Next term is equal to
\begin{equation}
\frac{d^2F(0)}{d^2t}
=-K_L(\mathcal{I})\star[Q,K]_L(\mathcal{I})+
[Q,K]_L(\mathcal{I})\star K_L(\mathcal{I})
=-[[Q,K],K]_L(\mathcal{I})=E_L(\mathcal{I}) 
\end{equation}
and consequently we obtain (Up to the second order, we
will see that  commutators of higher order  vanish.)
\begin{eqnarray}
 Q'(\Psi)=
 (Q-D)(\Psi)+\frac{1}{2}
 E_L(\mathcal{I})\star \Psi-
(-1)^{|\Psi|}\frac{1}{2} \Psi \star E_L(
\mathcal{I})=\nonumber \\
=(Q-D)(\Psi)-\frac{1}{2}\mathcal{I}\star E_R(\Psi)
-\frac{1}{2}E_L(\Psi)\star \mathcal{I}=
\left(Q-[Q,K]+\frac{1}{2}[[Q,K],K]\right)(\Psi) \ ,
\nonumber \\
\end{eqnarray}
where
\begin{eqnarray}
[[Q,K],K]=
\left[\frac{a}{\pi}\int_0^{\pi}
d\sigma 
\left\{c^0(\sigma)X'(\sigma)(\delta_N (\sigma,\pi)
-\delta_N(\sigma,0))\right\},
ia\int_0^{\pi}d\sigma'P(\sigma')\right]= \nonumber \\
=-\frac{ia^2}{\pi}\int_0^{\pi}d\sigma
\int_0^{\pi}d\sigma'\left[c^0(\sigma)\partial_{\sigma}
i\delta_D(\sigma',\sigma)(\delta_N(\sigma,\pi)-
\delta_N (\sigma,0))\right]=\nonumber \\
=-\frac{a^2}{\pi}\int_0^{\pi}d\sigma
\left[c^0(\sigma)(\delta_N(\sigma,\pi)-
\delta_N(\sigma,0))^2\right] \ .\nonumber \\
\end{eqnarray}
Now it is clear that commutators
of  higher order  vanish
and we obtain the final form of the shifted
BRST operator
\begin{eqnarray}\label{Q'n}
Q'=Q-\frac{a}{\pi}\int_0^{\pi}
d\sigma \left[\left(c^+\partial_+X-
c^-\partial_-X\right)\left(
\delta_N(\sigma,\pi)-
\delta_N (\sigma,0)\right)
\right] +\nonumber \\
+\frac{a^2}{4\pi}\int_0^{\pi}d\sigma
\left[(c^++c^-)(\delta_N(\sigma,\pi)-
\delta_N(\sigma,0))^2\right] \ .
\nonumber \\ 
\end{eqnarray}
We see the remarkable fact that 
the deformation of the BRST operator is
 the same as in the CFT description
\cite{SenSFT,Schomerus} in case of 
the matter sector, however we also see that there is
an additional term which is in fact singular
on the boundary and which contains the
ghost fields only. At present, we do not
completely understand reasons and meanings
of this term, it is possible that more
detailed analysis performed according
to the paper \cite{SenSFT} will explain
emergence of this term. We will return to 
this question in future work. 

However in spite of the fact that we have
such a unusual additional term, we can still
see that according to the general construction
given in the (\ref{second}) section after
performing field redefinition with the operator
$K$ the new BRST operator 
(\ref{Q'n}) is mapped to the original one and
the original string field $\Psi$ maps as follows
\begin{equation}
\tilde{\Psi}=e^{K}(\Psi)=\Psi \ 
\end{equation}
which is again consequence of the fact that any
string field does not carry momentum in the transverse
dimensions  and hence are invariant under the action
of $e^K$.

In this section we gave an example of the exact string
field theory solution corresponding to the 
translation of D0-brane in the transverse  space.
We have seen that this solution is in some sense
very simple  since the string field theory description
of D0-brane should not depend on the position
of D0-brane in the transverse space. On the other hand, as
was argued in \cite{SenM1,SenM2,SenSFT}
problem of marginal deformation is quiet nontrivial
in the context of the string field theory so that we believe
that our approach could be helpful in further study
of the string field theory. In the next section we give
the second example corresponding to the rotation
of the D0-brane position  in the transverse  space.
%%%%%%%%%%%%%%%%%%%%%%%%%%%%
%%%%Rotation of single D-brane %%%%%%%%
%%%%%%%%%%%%%%%%%%%%%%%%%%%%%
\section{Solution corresponding to
the rotation in the transverse space}\label{fifth}
Let us consider D0-brane that is localized
 in any  point in the transverse space
 with coordinates
$Y^I, I=1,\dots,25$ and consider solution of
the string field theory equation of motion
that  corresponds to the rotation of this 
D0-brane. As in previous example given in 
(\ref{third}) section  Dirichlet boundary conditions
explicitly break rotation invariance.
However we can still expect that the theory should
not depend on the position of D0-brane in the
transverse  space. 
Let us consider following transformation
of coordinates
\begin{equation}
\delta X^{I}=\omega^{I}_{J}X^{J} \ , 
\omega^I_J=const \ .
\end{equation}
It is easy to see that the world-sheet Lagrangian
is invariant under this transformation. Then
the generator of rotation symmetry is given
\cite{WittenSB} 
\begin{equation}
M^{IJ}=i\int_0^{\pi}d\sigma J^{0IJ}=
\frac{i}{4\pi\alpha'}\int_0^{\pi}d\sigma (
X^I\dot{X}^J-X^J\dot{X}^I) \ .
\end{equation}
Let us consider string field
\begin{equation}
\Phi_0=Q(\omega_{IJ}M^{IJ}_L(\mathcal{I}))=D_L(\mathcal{I}) ; \ 
D_L=[Q,\omega_{IJ}M^{IJ}]_L \ .
\end{equation}
It is clear that $\Phi_0$ solves the
 linearised equation of motion 
$Q(\Phi)=0$ thanks to the nilpotence of $Q$. As in previous case
the  new BRST operator can be written as
\begin{eqnarray}
Q'=Q-D \ , D=[Q,\omega_{IJ}M^{IJ}]=[Q_{bound},
\omega_{IJ}M^{IJ}] \ , \nonumber \\
Q_{bound}=-\frac{1}{\pi}\int_o^{\pi}
d\sigma c^0(\sigma)Y^K\eta_{KL}
X'^L(\sigma)(\delta_N (\sigma,\pi)-
\delta_N (\sigma,0) ) \ . \nonumber \\
\end{eqnarray}
where we have used manifest rotation invariance
of the bulk part of the BRST operator. We have also
included into bulk BRST operator the term proportional
$Y_IY^Ic^0(\sigma)(\delta_N(\sigma, \pi)-\delta_N(\sigma,0))^2$
that is manifestly invariant under rotation as well. 

We have to say a few words to the fact that
we have introduced boundary term $Q_{bound}$.
 As we argued in
the previous section we can perform field redefinition
and transform the BRST operator $Q+Q_{bound}$ to the
original one $Q$ and we have seen that the new string
field is the same as the original one, which was a
consequence of the fact that D0-brane physics in trivial
background does not depend on D0-brane position
in the transverse space. In the same way we could start with D0-brane
in the position $Y^I=0$ and consequently it would not
have a sense to speak about the solution of the string field equation
of motion corresponding to the rotation of D0-brane. 
However we would like to illustrate 
on this example an application of the  general
method given in (\ref{second})
section to generate solution of the equation of motion
hence we have introduced the  boundary term in the BRST operator
that reflects the fact that D0-brane is placed in
the position $Y^I$. We will perform more interesting
calculation in case of $N$ D0-branes in separate
paper. 
It is easy to perform calculation of 
the commutator with the result
\begin{equation}
[Q_{bound}, \omega_{IJ}M^{IJ}]=
-\frac{1}{\pi}\int_0^{\pi}d
\sigma c^0(\sigma)Y^I\omega_{IJ}X'^J(\sigma)
 (\delta_N(\sigma,\pi)-\delta_N (\sigma,0)) \ .
\end{equation}
As a result, the new BRST operator $Q'$ is
\begin{equation}
Q'=Q-D=Q_{bulk}
-\frac{1}{\pi}\int_0^{\pi}d
\sigma c^0(\sigma)Y^I(\eta_{IJ}+\omega_{IJ})
X'^J(\sigma)\left(
\delta_N(\sigma,\pi)-\delta_N (\sigma,0)\right) \ ,
\end{equation}
We see that our solution
 corresponds to the rotation in the transverse
space  to the new Dirichlet boundary conditions
\begin{equation}
Y'^K=Y^I(\delta_{I}^K
+\omega_{IM}\eta^{MK})=(\delta^K_I-
\eta^{KM}\omega_{MI})Y^I \ .
\end{equation}
It is a  simple exercise
to  generalize previous solution  to the
case of   finite 
rotation. Then we obtain
\begin{equation}
\Phi_0=e^{-K_L(\mathcal{I})}\star
Q\left
(e^{K_L(\mathcal{I})}\right) =
-\sum_{n=1}^{\infty}
\frac{1}{n!}[K,[K,\dots,[K,[K,Q]],\dots]_L
(\mathcal{I})=D_L(\mathcal{I}) \ 
\end{equation}
so that the new BRST operator is equal to 
\begin{eqnarray}
Q'=Q_{bulk}+Q_{bound}-D=Q_{bulk}+Q'_{bound} \nonumber \\
Q'_{bound}=-\frac{1}{\pi}\int_0^{\pi}d\sigma
c^0(\sigma)Y^I\Lambda_{IJ}X'^J(\sigma)
\left[\delta_N(\sigma,\pi)-\delta_N(\sigma,0)\right] , \nonumber \\
\Lambda_{IJ}=\sum_{n=0}^{\infty}\frac{1}{n!}
(\omega)^n_{IJ} , \
 \ , (\omega)^2_{IJ}=\omega_{IK}
\eta^{KL}\omega_{LJ} \ , \dots \ ,\nonumber \\ 
\end{eqnarray}
where we have used
\begin{eqnarray}
[K,[K,Q]]=[\frac{1}{\pi}\int_0^{\pi}d
\sigma c^0(\sigma)Y^I\omega_{IJ}X'^J(\sigma)
 (\delta_N(\sigma,\pi)-\delta_N(\sigma,0)),
 \omega_{KL}M^{KL}]=
\nonumber \\
=\frac{1}{\pi}\int_0^{\pi}d\sigma c^0(\sigma)
[\delta_N (\sigma,\pi)-\delta_N(\sigma,0)]
Y^I\omega_{IJ}\eta^{JL}\omega_{KL}
\partial_{\sigma}\int_0^{\pi}d\sigma' 
\delta_D (\sigma',\sigma)X^K(\sigma')
 =\nonumber \\ 
=\frac{1}{\pi}\int_0^{\pi}
d\sigma c^0(\sigma) Y^I\omega_{IJ}\eta^{JL}
\omega_{LK}X'^K(\sigma)
[\delta_N(\sigma,\pi)-\delta_N(\sigma,0)] \nonumber \\
\end{eqnarray}
so that
\begin{equation}
D=\frac{1}{\pi}\int_0^{\pi}d\sigma \sum_{n=1}
^{\infty}\frac{1}
{n!}Y^I(\omega^n)_{IJ}X'^J(\sigma)
[\delta_N (\sigma,\pi)-\delta_N (\sigma) ] \ .
\end{equation}
We see that the new boundary term $Q'_{Bound}$ 
corresponds precisely to the  situation
where D0-brane is placed in the points $Y'^I$.
Following our general discussion in (\ref{second}) section
we
can perform the field redefinition that maps the new
BRST operator $Q'$ to the original one $Q$ and
the original field is mapped to
\begin{equation}
\Psi'=e^K(\Psi) \ .
\end{equation} 
In this case however the new field $\Psi'$ will differ
from the field $\Psi$ since various components can
carry different tensor components and hence $K$
acts non trivially on these string states. Following
standard methods \cite{WittenSB} it is a  simple task
to obtain the new form of these fields. We will not
repeat this calculation here. 

%%%%%%%%%%%%%%%%%%%%%%%%%%%%%%%%%%%%%%%
%%%%Conclusion %%%%%%%%%%%%%%%%%%%%%%%%%%%%
%%%%%%%%%%%%%%%%%%%%%%%%%%%%%%%%%%%%%%%
%%%%%%%%%%%%%%%%%%%%
%%%%%%%%%%%%%%%%%%%%
\section{Conclusion}\label{sixth}
In this paper we have presented two examples
of the classical solution of the open
bosonic string field theory,
following our general discussion given in
paper \cite{KlusonGS}. We have seen that this
method gives an expected result in case of small
marginal deformation corresponding to translation
in transverse space to the world-volume of D0-brane,
however we have seen that in the case of finite
translation there is additional term that contains
only ghost contribution and is in fact singular
on the boundary of the world-sheet.
 At present we do not precisely
understand nature of this term. It is possible that it
can be avoided by careful CFT calculation, following
\cite{SenSFT}. We are going to make such a calculatin.
On the other hand, we have argued that with the appropriate
string field redefinition we can map the new BRST operator
to the original one and hence we can easily avoid
the problems with this term. 

To sum up, we hope
that our approach gives a new view on the  problem
of the marginal deformation in the open string field
theory.
We also believe that
in  more interesting examples, such as  a marginal
deformation of the configuration of $N$ D0-branes, our approach
could be very useful. We return to this problem in
our future work.

Of course, our approach is far more complete. As we claimed
above, it would be nice to prove detailed
analysis based on conformal field theory techniques,
following seminal paper \cite{SenSFT}.
Again we hope to return to this problem in our
future research.
\\
\\
{\bf Acknowledgment}
It is  my great pleasure to thank to  all people at the Department
of Theoretical Physics in Uppsala, especially to
Joseph Minahan, Ulf Danielsson and Konstantin Zarembo for
 useful discussions. 
I am also very graceful to
Ulf Lindstrom for his  support of  my research.
This work is partly supported 
by EU contract HPRN-CT-2000-00122.
This work is also supported by the Czech
Ministry of Education under Contract No.
143100006.

%%%%%%%%%%%%%%%%%%%%%%%%%%%%%%%%%
%%Apendix %%%%%%%%%%%%%%%%%%%%%%
%%%%%%%%%%%%%%%%%%%%%%%%%%%%%%%%
\section{Appendix}
In this appendix we will discuss some properties of
the delta function \footnote{For very nice discussion, see paper
\cite{Takahashi1}.}
\begin{equation}
\delta (\sigma,\sigma')_D=\frac{1}{\pi}
\sum_{n=-\infty}^{\infty}\sin n\sigma
\sin n \sigma'\ .
\end{equation}
that obeys Dirichlet boundary conditions 
\begin{equation}
\delta (\sigma,\pi)_D=\frac{1}{\pi}
\sum_{n=-\infty}^{\infty}\sin n\sigma
\sin n \pi =0  , \
\delta (\sigma,0)_D=0 \ .
\end{equation}
It is easy to see that for
 any function obeying Dirichlet boundary
condition 
\begin{equation}
f(\sigma)=\sum_{m=-\infty}^{\infty}
f_m \sin m \sigma \ 
\end{equation}
we  \  get
\begin{eqnarray}
\int_0^{\pi} d\sigma f(\sigma) \delta_D (\sigma,
\sigma')=\frac{1}{\pi}
\sum_{m,n}\int_0^{\pi}f_m\sin m\sigma sin n\sigma
\sin n\sigma'=\nonumber \\
=\frac{1}{\pi}
\sum_{m,n}\int_0^{\pi}d\sigma \frac{1}{2}\left[
\cos (m-n)\sigma-\cos (m+n)\sigma\right]
\sin n\sigma'
=\sum_{m=-\infty}^{\infty}f_m \sin m\sigma'=
f(\sigma') \ . \nonumber \\
\end{eqnarray}
For our purposes it is also very important following
formula
\begin{eqnarray}\label{dDN}
\int_0^{\pi}d\sigma' \partial_{\sigma}
\delta (\sigma,\sigma')_D=
\int_0^{\pi}d\sigma'\sum_n \frac{n}
{\pi} \cos n\sigma
\sin n \sigma'= \nonumber \\
= -\sum_n \frac{n}{\pi}
 \cos n \sigma \frac{1}{n}
(\cos (n\pi)-\cos (n 0))=
-(\delta_N(\sigma,\pi)-
\delta_N(\sigma,0)) \ ,\nonumber \\
\end{eqnarray}
where the  delta function
obeying Neumann boundary conditions is 
defined as
\begin{equation}
\delta (\sigma,\sigma')=\frac{1}{\pi}+
\frac{1}{\pi}\sum_{n=-\infty, n\neq 0}
^{\infty} \cos n 
\sigma \cos n \sigma' \ .
\end{equation}
%%%%%%%%%%%%%%%%%%%%%%%%%%%%%%%%%%%%%%
%%%%%%% Thebibligraphy %%%%%%%%%%%%%%%%%%%%%
%%%%%%%%%%%%%%%%%%%%%%%%%%%%%%%%%%%%%

\end{document}